 \documentclass{amos}

\usepackage{soul}
\usepackage{color}
\usepackage{hyperref}
\usepackage{graphicx}
\usepackage{tikz}
\usetikzlibrary{shapes.geometric}
\usetikzlibrary{arrows.meta,arrows}
\usetikzlibrary{positioning}

\title{On-chain Validation of Tracking Data Messages (TDM) Using Distributed Deep Learning on a Proof of Stake (PoS) Blockchain}
\author{\texttt{spaceprotocol.org}}
\author{Yasir Latif$^{*}$, Anirban Chowdhury, Samya Bagchi$^{*}$ - $^{*}$equal contribution\\Space Protocol\\
\texttt{\{yasir,anirban,samya\@spaceprotocol.org}}
\begin{document}
\maketitle
\begin{abstract}
\normalsize
    Trustless tracking of Resident Space Objects (RSOs) is crucial for Space Situational Awareness (SSA), especially during adverse situations. The importance of transparent SSA cannot be overstated, as it is vital for ensuring space safety and security. In an era where RSO location information can be easily manipulated, the risk of RSOs being used as weapons is a growing concern. The Tracking Data Message (TDM) is a standardized format for broadcasting RSO observations. However, the varying quality of observations from diverse sensors poses challenges to SSA reliability. While many countries operate space assets, relatively few have SSA capabilities, making it crucial to ensure the accuracy and reliability of the data. Current practices assume complete trust in the transmitting party, leaving SSA capabilities vulnerable to adversarial actions such as spoofing TDMs. This work introduces a trustless mechanism for TDM validation and verification using deep learning over blockchain. By leveraging the trustless nature of blockchain, our approach eliminates the need for a central authority, establishing consensus-based truth. We propose a state-of-the-art, transformer-based orbit propagator that outperforms traditional methods like SGP4, enabling cross-validation of multiple observations for a single RSO. This deep learning-based transformer model can be distributed over a blockchain, allowing interested parties to host a node that contains a part of the distributed deep learning model. Our system comprises decentralised observers and validators within a Proof of Stake (PoS) blockchain. Observers contribute TDM data along with a stake to ensure honesty, while validators run the propagation and validation algorithms. The system rewards observers for contributing verified TDMs and penalizes those submitting unverifiable data. This incentivizes the provision of accurate data and disincentivizes malicious activities. Furthermore, our blockchain-based approach opens a new, permissionless avenue for SSA data sourcing, validation, and recording. It encourages contributions from both traditional SSA operators and amateur observers such as those participating in SeeSat and SatNOGS, enhancing the diversity and reliability of RSO data. This comprehensive record of validated TDMs serves as a foundation for calculating RSO Detectability, Trackability, and Identifiability (DIT) scores on-chain, analogous to Space Sustainability Rating’s DIT module. We introduce this rating system as a decentralized application (dApp) running on the proposed blockchain. This system provides a transparent and immutable record of RSO sustainability scores, enhancing trust and accountability in space operations.\footnote{This is a live document. Please see the current version at \url{https://www.spaceprotocol.org/sda_chain_whitepaper.pdf}}

\end{abstract}

\section{Introduction}



The realm of space exploration and utilization has evolved significantly from being dominated by national space agencies to involving a diverse array of commercial, governmental, and international stakeholders. While advancing space technology and accessibility, this shift has introduced complex challenges related to space safety, sustainability, and security (S3). One of the most pressing issues is growing congestion in space, which is exacerbated by the proliferation of space debris. Current estimates indicate that there are over 128 million pieces of debris smaller than 1 cm, approximately 900,000 pieces between 1 cm and 10 cm, and approximately 34,000 pieces larger than 10 cm orbiting the Earth\footnote{\url{https://www.esa.int/Space_Safety/ESA_s_Space_Environment_Report_2023}}. These figures underscore not only the critical issue of space debris, but also the potential security threats to space assets.  Compounding these challenges is the lack of internationally accepted and enforced standardization of practices, which complicates coordination among multiple parties and can lead to misunderstandings and potential conflicts. To address these issues, there is a pressing need for a verifiable pre-launch supply chain to ensure that space assets are equipped with sufficient capabilities to maneuver and minimize fragmentation during incidents, thereby maximizing longevity.  Furthermore, decentralization of space sensing, storage, and computation is crucial to prevent any single entity from manipulating information with malicious intent. This decentralization enhances transparency and verifiability, which are essential for maintaining S3. Additionally, incentivizing resource sharing in space, including communication infrastructure and data sharing, is vital. Such a collaboration would optimize the in-orbit population, reduce the need for redundant systems, and mitigate the risk of space debris.  Addressing these dimensions through a concerted effort involving all space faring nations and entities is essential to ensure the long-term sustainability, safety, and security of outer space. Developing and implementing standards and frameworks that promote transparency and verifiability in structural integrity, behavioral practices, and collaborative endeavors will help safeguard the space environment for future generations.

In this work, we focus on the space domain awareness blockchain, the various entities that play pivotal roles in it, the incentivization mechanisms, and the services such a chain enables.


\subsection{Decentralization and its importance for space}

Decentralization plays a pivotal role in enhancing the security of applications in general for several reasons, fundamentally altering the way data is stored, managed, and protected. Decentralization, in the conext of space, is required as it:
\begin{itemize}
    \item \textbf{Eliminates the need for a single trusted party}: Data from many sources, verified and validated through consensus, makes a single source of truth obsolete.
    \item \textbf{Provides transparency and builds trust}: Decentralization provides a transparent and verifiable record of events and the sequence in which they occurred. This transparency helps build trust among participants, as actions within the network are traceable and auditable by all. 
    \item \textbf{Ensures data integrity and availability}: data is consistently available to users, is duplicated across the network, and the integrity of data is ensured through consensus algorithms, making unauthorized alterations easily detectable.
\end{itemize}


\subsection{The role of blockchain in S3}

To mitigate the lack of transparency and verifiability, we utilized the trustless and permissionless attributes of a blockchain. Specifically, we propose the deployment of three independent specialized blockchains to enhance S3. Each chain focuses on the domains of the manufacturing and supply chain, resident space object data behavioral dynamics, and on-orbit collaborative operations and services, respectively. By doing so, we can achieve separation of concerns, enabling the development of tailored governance and operational frameworks that can more effectively address the unique verification needs and challenges of each domain.  This tripartite blockchain framework optimizes data integrity, security, and operational efficiency across the key domains of S3 for Space, providing a scientific and professional strategy for addressing the challenges of the sector with a high degree of specificity and effectiveness. In the following section, we elaborate on the purpose and philosophy of each chain. 

\subsubsection{Blockchain to manage the satellite manufacturing supply chain}
Blockchain technology can significantly enhance the satellite manufacturing supply chain by providing a transparent and immutable record of each component's origin, quality, and compliance with the standards~\cite{SupplyChain}. This transparency allows manufacturers to quickly identify and select high-quality parts, reducing the likelihood of failure and extending the longevity of satellites. By ensuring that only verified and reliable components are used, the blockchain can help prevent premature satellite failures, thereby maximizing returns from space operations and contributing to a reduction in space debris.  Blockchain can also play a crucial role in reducing insurance fraud, in addition to improving the quality and reliability of satellite components. With a transparent and auditable record of all transactions and component histories, insurers can more accurately assess risks and validate claims, thus minimizing fraudulent activities. This enhanced traceability reduces the potential for insurance fraud and supports the overall sustainability of space operations by ensuring that only high-quality and compliant parts are used, thereby decreasing the likelihood of generating additional debris in space. Blockchain technology helps to create a safer and more sustainable space environment by fostering a more reliable and efficient supply chain.


\subsubsection{Space Domain Awareness Blockchain}

This blockchain specifically focuses on decentralized trustless sensing, storing, and processing data related to the behavior and dynamics of resident space objects, which include active satellites, defunct spacecraft, and space debris. At its core, this blockchain creates a decentralized and immutable ledger of RSO data sourced from a diverse network of space sensors. The need for such capabilities has already been felt in the community~\cite{BESTA, SISE}. The key advantage of this system is its decentralized nature, which prevents any single entity from monopolizing space sensing. Instead, it promotes collaborative heterogeneous tasking and sensing, in which data from various types of sensors contribute to a comprehensive and accurate picture of the space environment.  This collaborative approach enables the global calibration of sensors and facilitates data assimilation and cross-validation. For instance, observations from optical telescopes can be corroborated by radar measurements, enhancing the overall accuracy and reliability of RSO tracking. This multisource verification and validation significantly improves the trustworthiness of the data, which is crucial for collaborative uncorrelated track processing and RSO mining for space situational awareness.  The blockchain architecture incorporates decentralized storage solutions, ensuring that the sensed data remain immutable and accessible. This prevents data manipulation and ensures the long-term preservation of historical RSO behavior. Additionally, the system leverages decentralized computing to power a marketplace for decentralized applications (dApps) that can perform various analyses on RSO data. These dApps can offer services ranging from collision prediction to space traffic management, all of which operate on the foundation of trustworthy, blockchain-verified data.  By combining decentralized sensing, storage, and computation, the SDA Blockchain creates a robust ecosystem for space-data management. This system not only enhances the accuracy and reliability of space object tracking, but also democratizes access to critical space data, fostering innovation and collaboration in the space sector. 

In this work, we outline the architecture and processes that can enable a space domain awareness blockchain. 


\subsubsection{On-orbit Collaborative Operations and Services Blockchain}
On-orbit collaboration plays a crucial role in reducing redundancy and infrastructure burden in space operations. By enabling services such as maintenance, refueling, and upgrading directly in orbit, satellites can remain operational longer, reducing the need for frequent launches and minimizing resource consumption. This approach extends the lifespan of space assets and contributes to a reduction in space debris.  Data and infrastructure sharing are vital components of on-orbit collaboration. By sharing computational resources and platforms, satellites can perform complex tasks without requiring individual onboard systems, thereby reducing both weight and cost. This collaborative model is essential for a sustainable space economy, optimizing the use of limited resources, and minimizing the environmental impacts of space missions.  Moreover, a Bitcoin-like payment network is necessary to facilitate on-orbit operation and collaborative services. SpaceChain ~\cite{spacechain2023} is at the forefront of this innovation, integrating blockchain technology with satellite operations to enable secure, decentralized data sharing, and infrastructure use. By providing a transparent and tamper-proof environment, SpaceChain supports efficient and trustworthy collaboration in space, thereby enhancing operational efficiency and sustainability.

\begin{figure}
    \centering
    \includegraphics[width=0.6\linewidth]{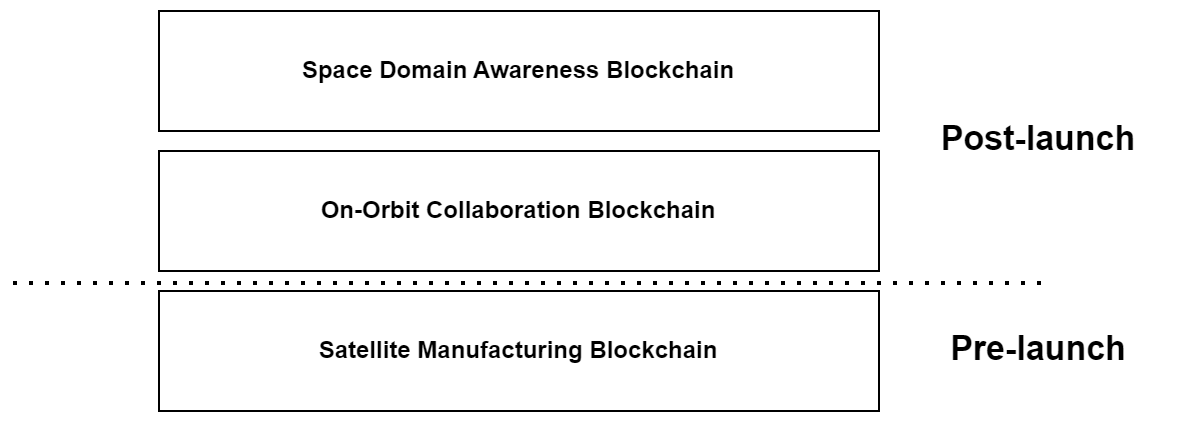}
    \caption{The three proposed chains for space safety, sustainability, and security.}
    \label{fig:distributed}
\end{figure}

\section{Resident Space Object (RSO) Tracking in Space Situational Awareness (SSA)}

Space Situational Awareness (SSA) is a critical aspect of space operations involving the collection, analysis, and tracking of detailed knowledge of resident space objects (RSOs) in the space environment. This includes monitoring the location, trajectory, and characteristics of satellites, space debris, and other objects in the Earth's orbit.

\subsection{Track Correlation and Matching}
The tracks are correlated and matched against known orbits through a process involving multiple steps. Initially, sensor data from various sources, such as ground-based telescopes and space-based surveillance systems, are collected and analyzed to identify and track objects in space. Data is then compared with a catalog of known orbits to determine whether the observed objects match any existing entries. Advanced algorithms and techniques, such as Multiple Hypothesis Tracking (MHT)~\cite{MHT}, are used to accurately track objects and resolve uncorrelated tracks (UCTs) and uncorrelated optical observations (UCOs).  

\subsection{Processing Uncorrelated and Ambiguous Tracks}

Uncorrelated and ambiguous tracks pose significant challenges to SSA. These tracks are processed using sophisticated methods to determine their origin and trajectory. For instance, the Adaptive Markov Inference Game Optimization (AMIGO)~\cite{AMIGO} SSA tool utilizes artificial intelligence and machine learning to analyze and predict the behavior of ambiguous tracks. However, a single company or entity may not have the capability to resolve all ambiguous tracks, owing to limitations in sensor coverage and data availability.  

\begin{figure}[t]
    \centering
    \includegraphics[width=0.9\linewidth]{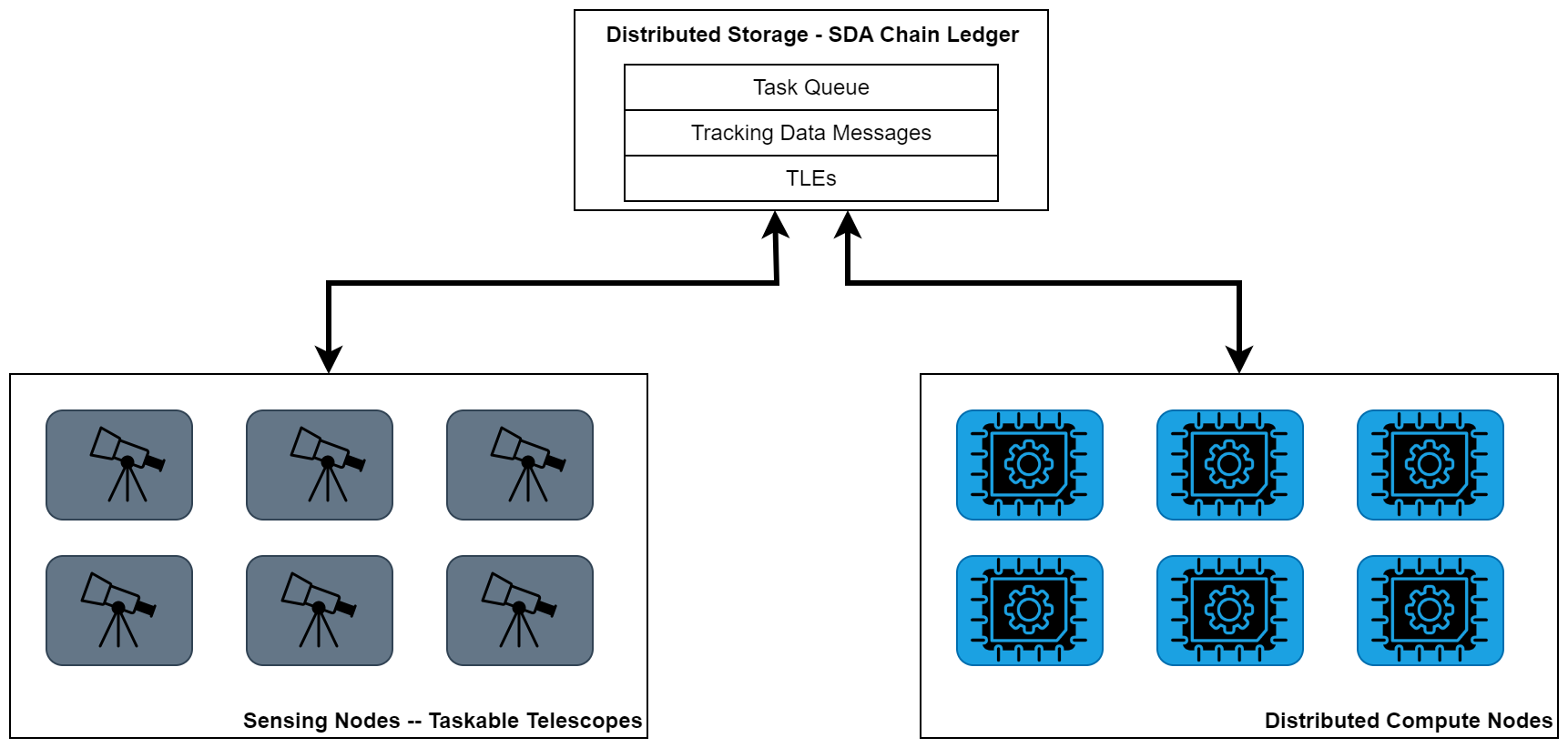}
    \caption{An overview of the SDA chain ecosystem consisting of a ledger (storage), decentralized sensing and compute nodes. See Sec. \ref{sec:bdb-ecosystem} for details about each component.}
    \label{fig:enter-label}
\end{figure}

\subsection{Collaborative Approach and Tracking Data Messages (TDM)}

A collaborative approach among multiple companies and organizations can significantly enhance the effectiveness of SSA. By sharing data and coordinating efforts, entities can leverage a broader range of sensors and resources to cover more of the sky at various depths. This collective effort can help to unravel the true story behind ambiguous tracks, providing a more comprehensive understanding of the space environment. Tracking Data Messages (TDMs) play a crucial role in this collaborative approach. TDMs are standardized messages used to share tracking data among different entities, facilitating the exchange of information and enhancing the accuracy of the SSA. However, the current system of sharing TDMs relies on centralized authorities and trust-based mechanisms, which can lead to vulnerabilities and inefficiencies.  

\subsection{Motivation for a Trustless Coordination System}

The limitations of the current system highlight the need for a trustless coordination system for telescope tasking and SSA data-processing as already pointed out in ~\cite{SNARE, SISE}. Such a system would enable secure, transparent, and decentralized data-sharing and coordination, overcoming the challenges of trust and centralization. Blockchain technology offers a promising solution for this problem. 

\subsection{Blockchain-Based Solution}

Blockchain technology provides a decentralized, immutable, and transparent platform for data-sharing and coordination. By utilizing blockchain, entities can securely share TDMs and coordinate telescope tasking without relying on centralized authorities or trust-based mechanisms. This approach ensures the integrity and accuracy of SSA data while also enhancing the efficiency and scalability of the system. In a blockchain-based system, TDMs can be encrypted and stored in a distributed ledger, ensuring that the data is secure and tamper-proof. Smart contracts can be used to automate the coordination of telescope tasking and data processing, eliminating the need for centralized authorities. This decentralized approach enables a more robust and resilient SSA system, capable of handling the complexities of space operations. The integration of blockchain technology into SSA systems offers a promising solution for enhancing the accuracy, efficiency, and security of space operations. By leveraging the benefits of decentralization, immutability, and transparency, blockchain-based systems can overcome the limitations of current SSA systems, thereby providing a more comprehensive and reliable understanding of the space environment.

\section{Space Domain Awareness (SDA) chain ecosystem}
\label{sec:bdb-ecosystem}

\begin{figure}
    \centering
    \includegraphics[width=0.9\linewidth]{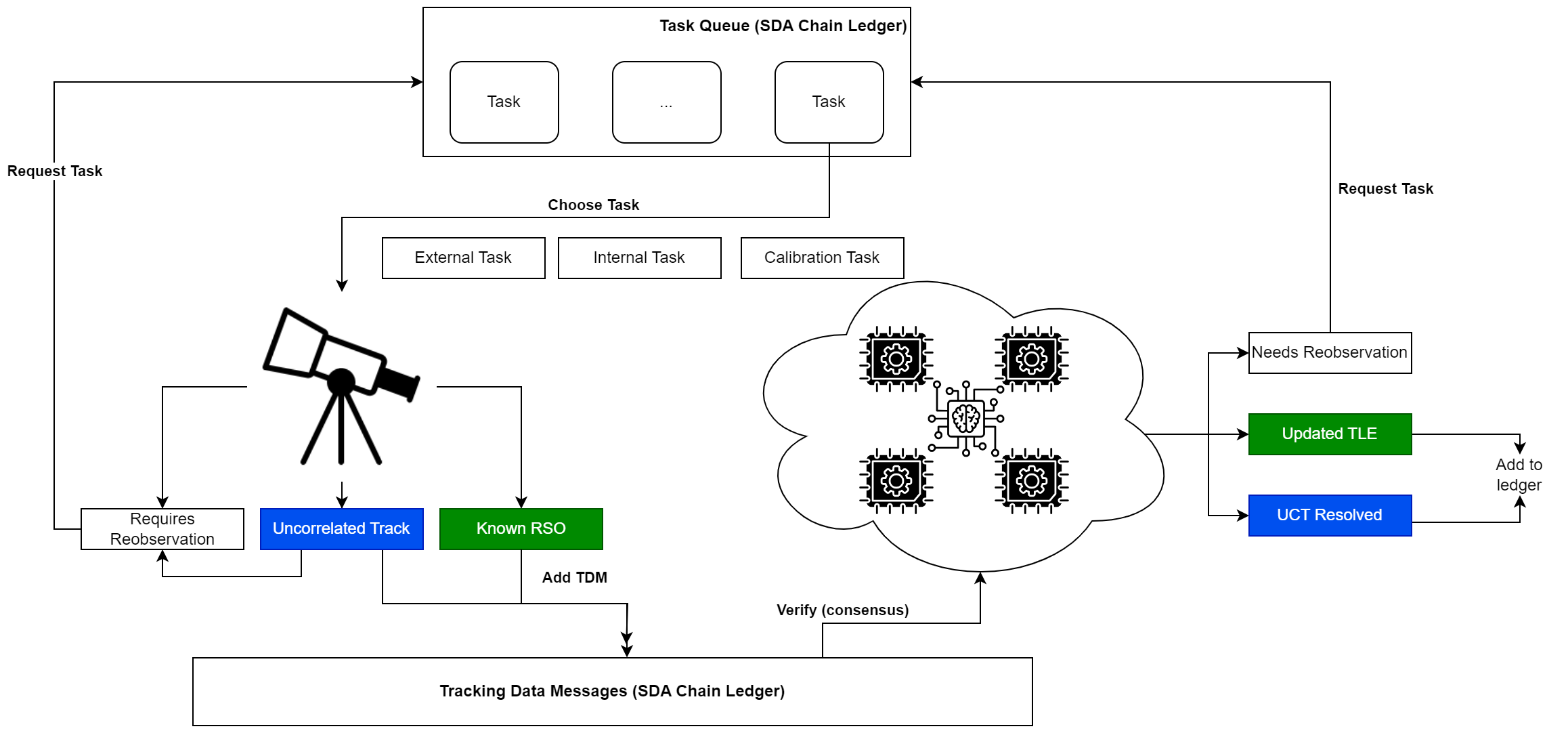}
    \caption{Operation of the SDA chain: Sensor tasking, data collection, decision making and recording the data onto the SDA chain. (See Sec. \ref{sec:bdb-ecosystem} for details.)}
    \label{fig:dataflow}
\end{figure}
The SDA chain ecosystem is designed to collect and store verifiable positioning data for Resident Space Objects (RSOs). This is achieved through a network of decentralized telescopes, which are tasked with observing specific RSOs based on a prioritized task queue maintained on the SDA chain, see Fig.~\ref{fig:dataflow} for an overview. In the following we describe the various components of the ecosystem.

\subsection{Task queue, prioritization, and  observations}
The task queue is populated with tasks from various sources including external customers who want to make particular observations, internal tasks generated from previous observations within the network, and calibration tasks, that enable sensors to perform dynamic self-calibration periodically. Tasks are prioritized based on factors, such as the age of the RSO, urgency (e.g., in the case of a break-up event), and classification status (e.g., Uncorrelated Tracks (UCTs)). 



Each sensor has algorithms to resolve observations against a catalog of known RSOs, which can be accessed from the SDA ledger or additional ``private'' RSO information. Sensors can retask an observation if they are not confident about the outcome, resolve the observation to a known object, or mark it as an uncorrelated track. Valid outcomes are registered onto the SDA ledger as Tracking Data Messages (TDMs) for network verification and validation.  

Requesters incentivize tasks by attaching a monetary benefit, which the sensors earn to make the requested observation, similar to the gas fees on the Ethereum chain. This payment mechanism also prevents network flooding caused by malicious users.  Similarly, when UCTs are resolved, that is, new RSOs are mined, miners or responsible sensor nodes are rewarded. This incentivizes the resolution of UCTs and contributes to the growth of the SDA chain ecosystem.  

\subsection{Verification and Validation of TDMs}  The TDMs generated by sensors are verified by compute nodes, which independently generate an Initial Orbit Determination (IOD) for the TDM, propagate it back in time to previous observations of the same object on the ledger, and assess its validity. Corrections to the object's IOD are carried out at this point to propose new Two-Line Elements (TLEs). Compute nodes that verify the observations are rewarded by the network using a lottery system.  
If the identity of an object is not known (UCT), additional track-to-track alignment algorithms are executed at the compute node to find the corresponding TDMs from the past. Successful resolutions are used to estimate a set of TLEs for the object. This is a computationally intensive task, and finding new associations is highly rewarded by the network, as it adds new information to the ecosystem. 

However, if an object is not resolved, an internal task request is generated to enable further observation of the UCT based on the Initial Orbit Determination (IOD) estimate. This leads to convergent behavior that prioritizes the resolution of UCTs and better orbit determination over successive sensor observations.  


\subsection{Sensor self-calibration and supervised learning}  Calibration satellites and services like the International Laser Ranging System (ILRS)~\cite{ILRS} provide precise ephemeris information, which can be used as a supervision to improve the quality of Two Line Elements (TLEs). The improved TLEs, along with the precise ephemeris information, are used to train a supervised propagation model, which is then used to predict the position of RSOs for validation. This approach ensures that the TLEs generated by the SDA chain ecosystem are both accurate and reliable.  


\section{Distributed Learning for Propagation}  Compute nodes require an orbit propagation algorithm to predict the position of RSOs for validation. Given the vast amount of observations and the ability to make frequent observations of the same object using decentralized sensing, we employ distributed learning to train a supervised propagation model. This model is trained using precise ephemeris data from services such as the International Laser Ranging Service (ILRS)~\cite{ILRS} and the Global Navigation Satellite System (GNSS)~\cite{GNSS}, ensuring that the TLEs generated by the SDA ecosystem are accurate and reliable.  Distributed Federated Learning for TDM verification  


\subsection{Challenges in IOD estimation and orbit propagation} 

A critical aspect of the SDA ecosystem is its ability to correlate and verify Tracking Data Message (TDM) observations across time. While tasking can play an important role by providing additional observations, sensor tasking itself requires accurate Initial Orbit Determination (IOD) to quantify the orbit of the observed RSO and enable observation using a different sensor.
 Correlating observations with known RSOs is a well-understood problem. However, in the case of Uncorrelated Tracks (UCTs), the problem is more complex as limited observations lead to ambiguities in the IOD process. Further errors are introduced by orbit propagation algorithms, such as SGP4~\cite{sgp4}, which is the default standard orbit propagation algorithm. Such accumulated errors lead to imprecise predictions of the target's visibility for the other sensors in the network. 


In this work, we assume that IOD algorithms work sufficiently well, and the main source of error is the orbit propagation algorithm. Constant addition of data to the chain enables opportunities to learn the characteristics of the space atmosphere from on-chain data. We utilize the advances made in Blockchain Federated Learning (BCFL)~\cite{li2022blockchain,qu2022blockchain} to incentivize nodes and collaboratively update a global orbit propagation model.  

\subsection{Federated Learning}  Federated Learning (FL) is a training paradigm that allows participating nodes to collaboratively train a model without sharing their data with one another or with a third party. In the case of the SDA chain, where data is accessible to all participants, using FL allows external parties that might not be actively engaged in the sensing aspects of the chain to still participate in updating the collaborative model. This enables them to keep sensitive data private and still contribute to updating the on-chain orbit propagation model. 

\subsubsection{Decentralized Federated Learning Mechanism}  
\begin{figure}[h]
    \centering
    \includegraphics[width=0.75\linewidth]{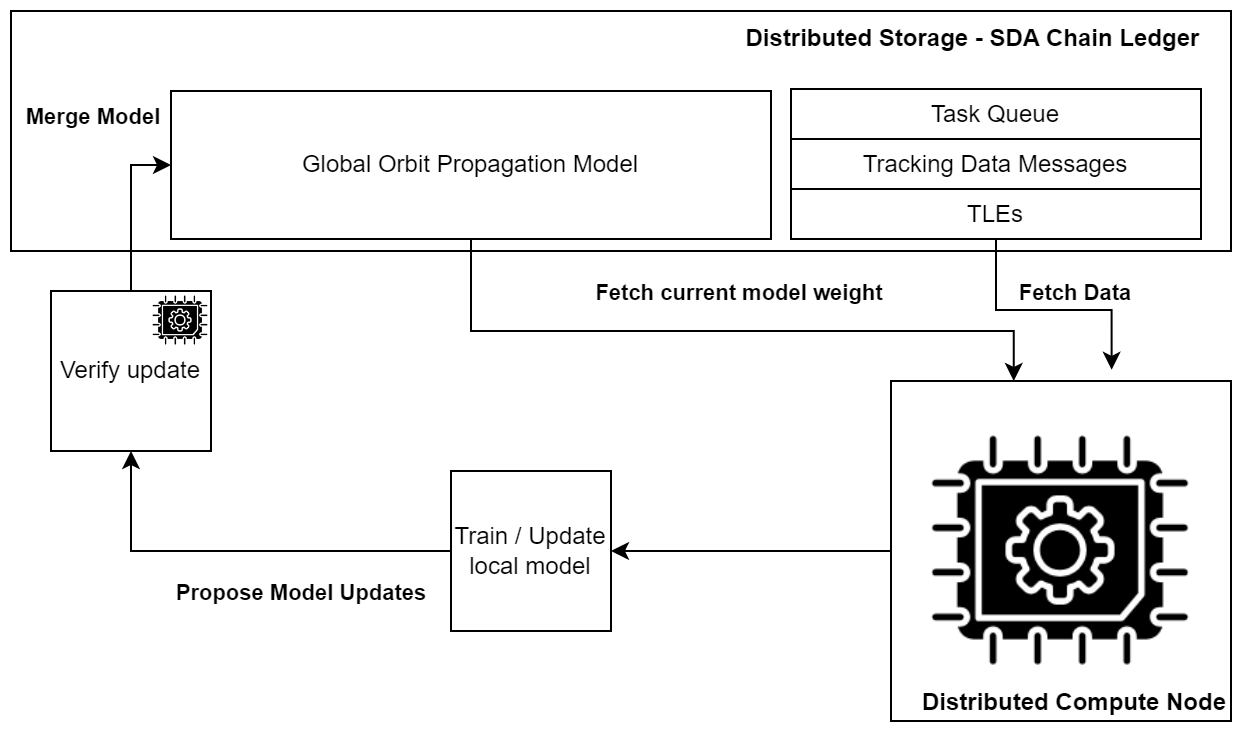}
    \caption{A schema of how the global propagation model is maintained and updated within the SDA chain ecosystem: compute nodes fetch data and current model from the network, and use the new data to refine the model parameters. These parameters are broadcast to other compute nodes, who can verify the capabilities of the new model since the have access to the corresponding training data in the network. Once the update has been verified by a set of participanting nodes, the global model is updated by merging the new weights.}
    \label{fig:bcfl}
\end{figure}

The decentralized federated learning mechanism is illustrated in Fig. \ref{fig:bcfl}. Compute nodes retrieve data from the chain along with the current model and run a training algorithm locally to compute updated model weights. Multiple observations of the same RSO over time combined with precise ephemeris from ILRS~\cite{ILRS} and GNSS~\cite{GNSS} provide supervision for training. The updated model must be verified by other compute nodes in the ecosystem, which use Two-Line Elements (TLEs) present on-chain to propagate RSOs and cross-verify the validity of the model. A voting scheme is employed to select the best model that is merged with the global model maintained on the chain. This updated model is then used by other nodes to verify and propagate TDMs.  For further details about the mechanism of the federated learning used in this work, we refer the reader to the work in \cite{10.1145/3422337.3447837}, which presents a flexible framework for BCFL and allows the use of both permissioned and public blockchains using a proof of stake based reward mechanism.

\section{Services as distributed Applications on the SDA chain ecosystem}

Verified data gathered on the SDA chain ecosystem enables a wide range of applications. Distributed applications (dApps) interact with the ecosystem to provide services that are crucial in every aspect of a satellite's lifetime. We highlight some of these applications below and how data on SDA chain can provide value:

\textbf{Space Sustainability Ratings} Using data on-chain, metrics for space sustainability such as L-DIT\cite{LDIT} and SSR\cite{SSR} can be computed in near real-time to ensure a better perspective of each player's response to space sustainability guidelines.    

\textbf{Launch Window Optimization} involves finding the optimal time frame for launching a spacecraft to ensure successful mission execution, considering factors such as trajectory planning and resource constraints~\cite{ip2022overview}. SDA chain can provide accurate RSO position information to plan a safe launch.  

\textbf{Orbit Planning} focuses on determining the best orbit for a spacecraft to achieve its mission objectives while avoiding collisions and interference with other space objects. SDA chain  provides data to analyze orbital densities. 

\textbf{Conjunction Analysis}~\cite{balch2019satellite,george2011comparison,singh2020new} involves assessing the risk of collisions between space objects by analyzing their orbits and predicting potentially close approaches. Accurate RSO data on the SDA chain ecosystem can provide tighter bounds on the uncertainty ellipses associated with the conjunction predictions.

\textbf{Maneuver Planning and Detection} involves designing and optimizing the maneuvers required for a spacecraft to reach its desired orbit or position, considering factors such as fuel efficiency and safety. SDA chain can provide accurate information about the positions of nearby RSOs.  Similar to Maneuver detection~\cite{kelecy2007satellite}, on-chain TLEs can be used.

\textbf{Launch Detection} involves identifying and tracking the launch of new space objects to maintain situational awareness and ensure safety in space operations. Object discovery and track association and integral parts of SDA chain ecosystem and enable early launch detection

\begin{figure}[h]
    \centering
    \includegraphics[width=0.5\linewidth]{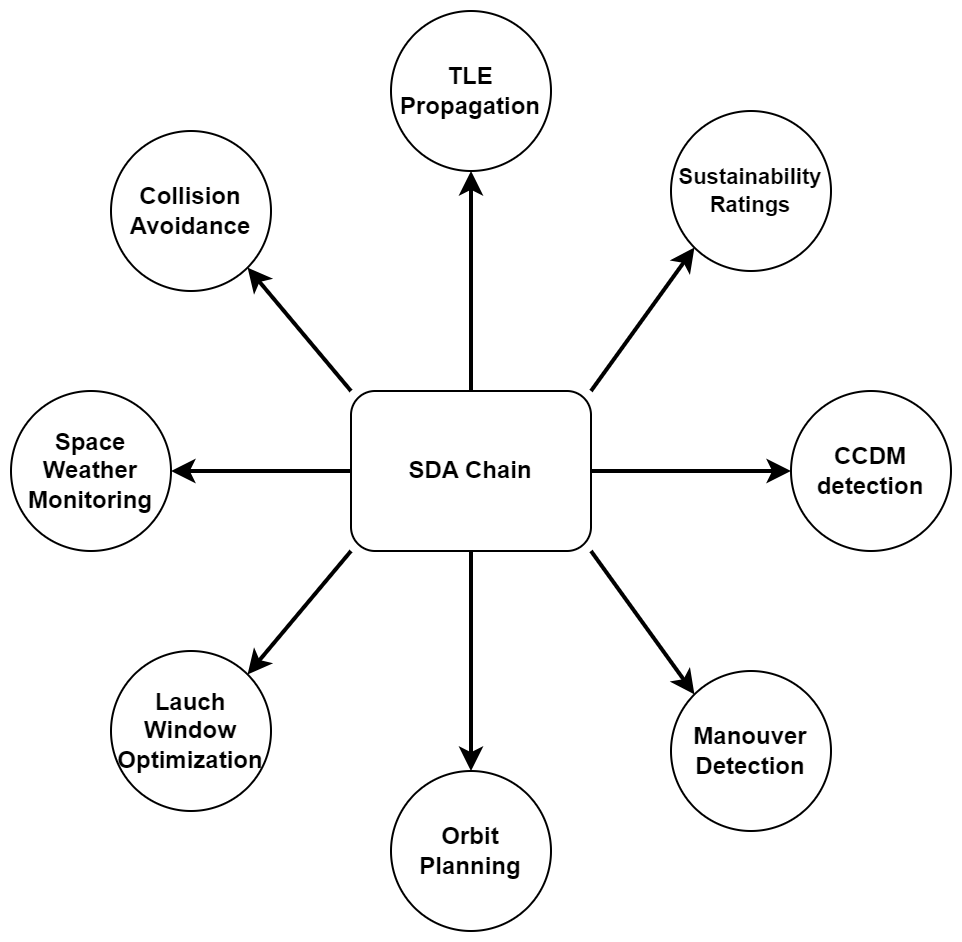}
    \caption{Caption}
    \label{fig:enter-label}
\end{figure}

\section{Conclusions and predictions for the future}
Even under peaceful conditions, the unsustainable practices prevalent within the current space operations paradigm are progressively exacerbating the space environment, rendering it increasingly hazardous for future generations. Threats to S3 are further exacerbated during coordination efforts among multiple players, where the lack of standardized communication and operational protocols can lead to misunderstandings and potential conflicts.

The challenges of enforcing S3 regulations across international borders pose additional challenges~\cite{mccormick2013space}. Given the global nature of space activities, unilateral or region-specific measures often fall short in addressing the collective needs for S3. This situation necessitates a global consensus and cooperation, which, in turn, requires a foundation built on transparency and verifiability. 

We believe that in order to uphold the principles of S3 effectively, a multi-faceted approach focusing on transparency and verifiability across three major fronts is crucial. 
\textbf{a)} Firstly, there is a need for enhanced transparency regarding the structural and physical properties and capabilities of space assets. Ensuring that satellites and other space assets are equipped with technologies to avoid collisions, generate fewer fragments in the event of an incident, and improve their detectability is paramount. 
\textbf{b)} Secondly, the behavioral dynamics of space assets, including satellite location sharing, coordination mechanisms, maneuvering strategies, and post-mission disposal plans, must be openly shared and verified. This transparency will facilitate the prediction and prevention of potential collisions and other hazardous incidents. 
\textbf{c)} Lastly, the collaborative nature of space operations must be encouraged and strengthened. By fostering an environment where infrastructure or software services can be shared among space assets, the overall in-orbit population can be optimized, reducing the need for redundant systems and thereby mitigating the risk of space debris.

It goes without saying that addressing these dimensions through a concerted effort involving all space-faring nations and entities is crucial for the long-term sustainability, safety, and security of outer space. The development and implementation of standards and frameworks that promote transparency and verifiability in structural integrity, behavioral practices, and collaborative endeavors represent a forward-looking strategy to safeguard the space environment for future generations.

\section{References}

\bibliographystyle{plain}

\begingroup
\renewcommand{\section}[2]{}%
\bibliography{references}
\endgroup

\end{document}